\documentclass[aps,twocolumn,showpacs]{revtex4}
\usepackage{graphicx}
\usepackage{bm}
\usepackage{amsmath}
\usepackage{amssymb}
\begin{document}

\title{Chiral Topological Superconductor From the Quantum Hall State}

\author{Xiao-Liang Qi$^{1,2}$, Taylor L. Hughes$^{1,3}$ and Shou-Cheng Zhang$^{1}$}
\affiliation{$^{1}$Department of Physics, Stanford University, Stanford,
    CA 94305, USA}
\affiliation{$^2$Microsoft Research, Station Q, Elings
Hall, University of California, Santa Barbara, CA 93106, USA}
\affiliation{$^{3}$Department of Physics, University of Illinois, 1110 West Green St, Urbana 61801, USA.}

\begin{abstract}
The chiral topological superconductor in two dimensions has a full
pairing gap in the bulk and a single chiral Majorana state at
the edge. The vortex of the chiral superconducting state carries a
Majorana zero mode which is responsible for the non-abelian statistics of the vortices.
Despite intensive searches, this novel superconducting
state has not yet been identified in nature. In this paper, we
consider a quantum Hall or a quantum anomalous Hall state near
the plateau transition, and in proximity to a fully gapped $s$-wave
superconductor. We show that this hybrid system realizes the
long sought after chiral superconductor state, and propose
several experimental methods for its observation.
\end{abstract}


\maketitle
There are two basic types of topological states in two dimensions (2D)
which break the time reversal symmetry $T$. The first is the quantum Hall (QH) state which
has a full gap in the bulk and gapless chiral  modes at the
edge. The integer number of the chiral edge modes $N$ is a topological invariant
which can be directly related to the bulk topological invariant of
the QH state\cite{thouless1982}. The quantum Hall state is realized
in the presence of a large external magnetic field, however, the
quantized Hall conductance can in principle also be realized in
topological insulators which break the $T$ symmetry\cite{Haldane1988,qi2005}.
More recently, realistic proposals suggest that doping 2D topological
insulators such as ${\rm HgTe}$ and ${\rm Bi_2Te_3}$ with magnetic dopants\cite{liu2009C,yu2010},
can result in the so-called quantum anomalous
Hall (QAH) insulator without an external magnetic field. The second state is the chiral
superconductor which has a full pairing gap in the 2D bulk, and ${\cal N}$
gapless chiral Majorana fermions\cite{wilczek2009} at the edge. The case of a chiral
superconductor with ${\cal N}=1$ is most interesting. The edge state
has half the degrees of freedom of an $N=1$ QH or QAH state, therefore,
this is the minimal topological state in 2D. The vortex of such a
chiral topological superconductor (TSC) carries a single Majorana zero mode,
\cite{volovik1999} giving rise to the non-abelian statistics\cite{read2000,ivanov2001}
which could provide a platform for topological quantum computing\cite{nayak2008}.
The simplest model for an ${\cal N}=1$ chiral TSC
is realized in the $p_{x}+ip_{y}$ pairing state of spinless fermions\cite{read2000}.
A spinful version of the chiral superconductor has been
predicted to exist in Sr$_2$RuO$_{4}$\cite{mackenzie2003}, however, the
experimental situation is far from definitive. Recently, several new proposals of realizing Majorana fermion state with conventional superconductivity have been investigated by making use of strong spin-orbital coupling\cite{fu2008,sau2010}.

In this letter we propose a general and intrinsic relation between
QH states and the chiral TSC state, which
leads to a new method to generate a chiral TSC from a QH or a QAH parent state. When a QH state
is coupled to a conventional s-wave superconductor through
the proximity effect, the topological phase transition between phases with
trivial and nontrivial Hall conductance is in general split into two
transitions, between which there is {\it always} a chiral TSC phase. Compared to conventional QH systems,
the QAH system can realize the QH state and the
topological phase transition without a large external magnetic
field, which makes the proximity effect to a superconductor much
easier to realize. Physically, our proposal is based on the
observation that the QAH system with $N$ chiral edge modes in proximity with a conventional
superconductor is already a chiral TSC with
an even number ${\cal N}=2N$ of chiral Majorana edge modes. Since the
degeneracy among these chiral Majorana modes is lifted by the proximity
to the superconductor, the transition from the QAH insulator to a topologically trivial
insulator must generically pass through a chiral TSC
phase with an odd number ${\cal N}$ of chiral Majorana edge modes. This
is the interesting state with non-Abelian statistics. 

\begin{figure}[t!]
\includegraphics[width=0.35\textwidth] {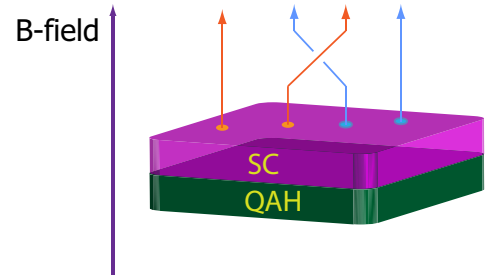}
\caption{Our proposed hybrid device consists of a QAH insulator
layer and a fully gapped superconductor layer on top. When the
QAH insulator is close to the topological quantum phase transition,
the proximity effect to the superconductor generically induces
a chiral topological superconductor phase with an odd number of chiral Majorana edge modes. In the presence of a magnetic
field ${\bf B}$, vortices carry Majorana zero modes with non-abelian
statistics. The QAH insulator could also be replaced by a QH state near
the plateau transition. } \label{fig:device}
\end{figure}

{\bf The QAH insulator} Our proposal works both for a QH state near the
plateau transition and a QAH state near the topological phase transition
to a trivial insulator. For definiteness, we focus on the QAH case in
most of our Letter, and comment on the generality and the QH state
near the end. We take the simplest model QAH Hamiltonian realized with low-energy states
near the $\Gamma$ point:\cite{qi2005}
\begin{eqnarray}
H_{QAH} & = &\sum_{\bf p}\psi_{\bf p}^\dagger h_{QAH}(\bf p)\psi_{\bf p} \\ \nonumber
h_{QAH}(\bf p) & = &\left(\begin{array}{cc}m(p) & A(p_{x}- i p_{y})\\
A(p_{x}+i p_y) & -m(p)\end{array}\right)\label{Hqah}
\end{eqnarray}
where $m(p)=m+Bp^2$, $A,B,m$ are material parameters and
$\psi_{\bf p}=\left(c_{p\uparrow}\;\;\;c_{p\downarrow}\right).$
For $B=0$, this Hamiltonian is exactly the massive Dirac Hamiltonian
in $2+1$ dimensions, however, the presence of the $B$ term in the QAH
Hamiltonian is crucial for determining the topological properties.
In the following we will take $B>0$. The bulk energy spectrum
$E_{\pm}(p)=\pm\sqrt{A^2p^2+(m+Bp^2)^2}$ is gapped as long as $m\neq
0.$ The Hall conductance of any gapped system is quantized, {\it i.e.}
$\sigma_H=Ne^2/h$, where $N$ is the first Chern number in momentum
space defined by\cite{thouless1982,qi2005}
\begin{eqnarray}
N=\frac1{2\pi}\sum_{E_n<0}\int
d^2p\left(\partial_xa^{nn}_y-\partial_ya^{nn}_x\right)\label{Chern}
\end{eqnarray}
with $a^{nn}_i=-i\left\langle n,{\bf p}\right|\partial/\partial
k_i\left|n,{\bf p}\right\rangle$ the Berry phase connection in
momentum space, and $n$ the band index. For the specific model
(\ref{Hqah}), the Hamiltonian can be rewritten as $h_{QAH}=
\sum_ad_a({\bf p})\sigma_a$ with
$\sigma_a$ the Pauli matrices, and $d_a({\bf p})=(Ap_x,Ap_y,m(p))$.
We see that the $m(p)$ term generally breaks the $T$ symmetry.
The Chern number for Hamiltonians of this form has a simple expression\cite{qi2005}
\begin{eqnarray}
N=\frac{1}{8\pi^2}\int d^2 p \epsilon^{abc}\hat{d_{a}}\frac{\partial
\hat{d_b}}{\partial p_x}\frac{\partial \hat{d_c}}{\partial
p_y}\label{windingnumber}
\end{eqnarray}
where the unit vector $\hat{d}_a({\bf p})=d_a({\bf p})/\sqrt{\sum
d_a^2({\bf p})}$. According to Eq. (\ref{windingnumber}), the Hall conductance is
determined by the winding number of the unit vector $\hat{\bf
d}({\bf p})$ in momentum space. It is straightforward to see
that for $m<0$ $\hat{\bf d}({\bf p})$ has a Skyrmion configuration
with $N=1$, while for $m>0$ the winding number is trivial $N=0$\cite{qi2005}. The
point $m=0$ is a quantum critical point between a trivial insulator
and a QAH insulator. The QAH phase with Chern number $N$ has $N$
chiral edge states. In the simplest case of $N=1$, the edge
state is described by the effective one-dimensional Hamiltonian
$H_{edge}=\sum_{p}vp\eta_{p}^\dagger \eta_{p}$, where $\eta_{p}^\dagger$
and $\eta_{p}$ are the creation and annihilation operators of the
chiral edge fermion.

\begin{figure}[t!]
\begin{center}
\includegraphics[width=0.45\textwidth] {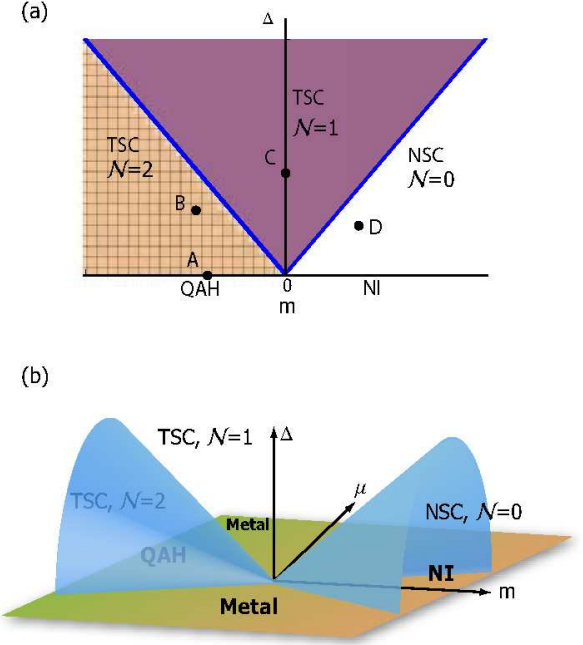}
\end{center}
\caption{(a) Phase diagram of the QAH-SC hybrid system for $\mu=0$. The x-axis
labels the mass parameter $m$ and the y-axis labels the magnitude of $\Delta.$
Integers ${\cal N}$ label the Chern number of the superconductor, which is equal to the number
of chiral Majorana edge modes. (b) Phase diagram for
finite $\mu$, shown only for $\Delta\geq 0$.
Phases QAH, NI and Metal (labeled in bold) are well-defined only in the $\Delta=0$ plane. } \label{MDphase}
\end{figure}

{\bf Phase diagram of QAH-SC system}
In proximity to an $s$-wave superconductor, a
finite pairing potential can be induced in the QAH state. The
BdG Hamiltonian for the proximity coupled QAH state is
\begin{eqnarray}
H_{BdG}&=&\nonumber\\
&\frac{1}{2}&\sum_{\bf p}\Psi^{\dagger}_{\bf
p}\left(\begin{array}{cc}h_{QAH}({\bf p})-\mu & i\Delta \sigma^{y}
\\-i\Delta^{*}\sigma^{y} &
-h^{*}_{QAH}(-{\bf p})+\mu\end{array}\right)\Psi_{\bf
p}\nonumber\\ \label{Hbdg}\end{eqnarray}
\noindent where $\Psi_{p}=\left(c_{p\uparrow}\;\;
c_{p\downarrow}\;\; c^{\dagger}_{-p\uparrow}\;\;
c^{\dagger}_{-p\downarrow}\right)^{T}$.  Just as in the QAH case, the
superconductor Hamiltonian (\ref{Hbdg}) can be classified by the
Chern number ${\cal N}$ defined in Eq. (\ref{Chern}) in momentum space. However, there are
two key differences from the QAH case. First, for the
superconductor Hamiltonian, the Chern number ${\cal N}$ does not physically correspond to
a quantized Hall conductance because \emph{charge} is not conserved. A superconductor with Chern
number ${\cal N}$ has ${\cal N}$ edge states, similar to the QAH system, but the number of edge
states is counted in the basis of chiral Majorana fermions, which are their own anti-particles. For
example, the edge state of an ${\cal N}=1$ TSC state is
described by $H_{edge}=\sum_{p\geq 0} vp\gamma_{-p}\gamma_p$ where the
chiral Majorana fermion operators $\gamma_p$ satisfy
$\gamma_{-p}=\gamma_p^\dagger,~\left\{\gamma_p,\gamma_{q}\right\}=\delta_{p+q}$.
Second, a superconductor has quantized vortices. A
TSC with \emph{odd} Chern number ${\cal N}$ generically has a Majorana zero mode in
the vortex core\cite{read2000}, which is described by a Majorana
operator $\gamma_0$ satisfying $\left[\gamma_0,H_{BdG}\right]=0$ and
$\gamma_0=\gamma_0^\dagger$. The Majorana zero mode is protected
topologically because the energy spectrum of the BdG Hamiltonian is
always symmetric around zero energy. The presence of the Majorana fermion is
essential for topological quantum computing applications\cite{nayak2008}.

We will first study the phase diagram of the system (\ref{Hbdg}) for
$\mu=0$. The bulk quasi-particle spectrum is $E_{\pm}({\bf
p})=\pm\sqrt{A^2p^2+(\Delta\pm m(p))^2}.$ Since topological invariants cannot change
 without closing the bulk gap, the phase diagram can be
determined by first finding the phase boundaries which are gapless
regions in the $(m,\Delta)$-plane, and then calculating the Chern number
of the gapped phases. For this
model the critical lines are determined by $\vert \Delta \pm m\vert
=0$, which leads to the phase diagram as shown in Fig. \ref{MDphase}
(a). (Only the region  $\Delta\geq 0$ is shown.) As expected, the
phase boundary reduces to the critical point $m=0$ between the QAH phase
and a trivial, or normal insulator (NI) phase in the limit $\Delta=0$. The point
$m=0$ is a multi-critical point in this phase diagram. For $m>0$ and
$\vert m\vert>\vert\Delta\vert$ the system is adiabatically
connected to a trivial insulator state with a full gap and no edge
state, so  it must be a trivial superconductor phase. For $m<0$
and $\vert m\vert>\vert\Delta\vert$ the system is in a non-trivial
TSC state which is adiabatically connected to
the QAH state in the $\Delta=0$ limit. The Chern number of this
phase can be determined by the $\Delta=0$ limit, in which case the
off-diagonal terms of the BdG Hamiltonian (\ref{Hbdg}) vanish. In this limit the eigenstates
of $H_{BdG}$ are determined by the eigenstates
of the QAH Hamiltonian $h_{QAH}({\bf p})$ (Eq.
(\ref{Hqah})). It is
straightforward to check from Eq. (\ref{Chern}) that the Chern
number $N_p$ and $N_h$ for the particle and the hole states are both
equal to that of the QAH system. In the $m<0$ phase we have
$N_p=N_h=1$, leading to the total Chern number ${\cal N}=N_p+N_h=2$.

Similarly, the Chern number of the superconductor phase emerging at
finite $\Delta$ can be determined by studying the special limit of
$m=0$ (point C in Fig. \ref{MDphase} (a)). In this limit, the
Hamiltonian (\ref{Hbdg}) can be block diagonalized by a basis
transformation into the following form:
\begin{eqnarray}
H_{BdG}({\bf p})&=&\frac12\sum_{\bf p}\tilde{\Psi}_{\bf
p}^\dagger\left(\begin{array}{cc}h_+({\bf p})&\\&-h_-(-{\bf
p})^*\end{array}\right)\tilde{\Psi}_{\bf p}\nonumber\\
\text{with}~h_{\pm}({\bf
p})&=&\left(\begin{array}{cc}\pm\left|\Delta\right|+Bp^2&A(p_x-ip_y)\\A(p_x+ip_y)&-\left(\pm\left|\Delta\right|+Bp^2\right)\end{array}\right).
\end{eqnarray}
Thus we see that the Hamiltonian is equivalent to two copies of the QAH
Hamiltonian (\ref{Hqah}) but with opposite mass parameters
$m=\pm\left|\Delta\right|$. The Chern number of $h_+({\bf p})$ is
trivial and that of $h_-({\bf p})$ is $N=1$, so that the total Chern
number of this TSC state is ${\cal N}=1$. In other words,
we have proven that a TSC phase with odd
Chern number ${\cal N}=1$ emerges in the neighborhood of the quantum
critical point between the QAH insulator and trivial insulator phases.

\begin{figure}[t!]
\includegraphics[width=0.45\textwidth] {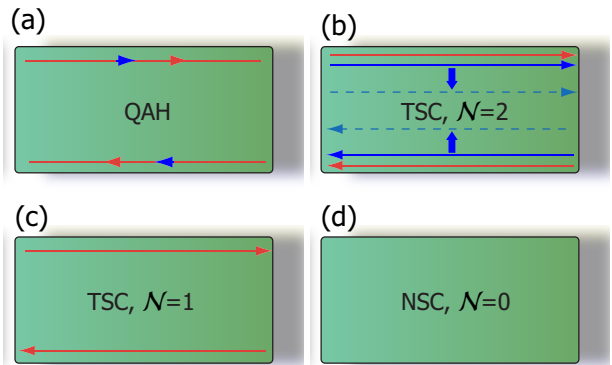}
\caption{Evolution of the edge states. The
four panels (a) (b) (c) (d) correspond to the edge state
configuration of points A, B, C, D in Fig. 2 (a), respectively. (a) and (b)
show that a single chiral edge mode of the QAH state can be decomposed into
two chiral Majorana edge modes of the TSC state. Proximity coupling to
the SC state lifts the degeneracy, and one pair of the chiral Majorana states
can be annihilated in the bulk, giving rise to a chiral TSC state with a
single chiral Majorana edge mode as shown in (c). Further changes of the
parameters can cause a phase transition into the trivial or normal SC (NSC)
state as shown in (d).}
\label{fig:edgeschematic}
\end{figure}

Next, we consider $\mu\neq 0$ in the Hamiltonian (\ref{Hbdg}), which
corresponds to proximity induced superconductivity in a {\em doped}
QAH system. Practically, the proximity effect is expected to be
stronger in this case due to the finite density of states at the
Fermi level. Similar to the $\mu=0$ case, we determine the phase
boundaries by the gapless regions in the energy spectrum, which leads to the
following condition 
\begin{eqnarray}
\Delta^2+\mu^2=m^2\nonumber .
\end{eqnarray}\noindent
The entire phase diagram in the $(m,\mu,\Delta)$ space is shown in
Fig. \ref{MDphase} (b). Except for the metallic phase in the $\Delta=0$
plane with $|\mu|>m$, and the phase boundaries, there are three
gapped phases. The Chern number of each phase can be determined by
its adiabatic connection to the $\mu=0$ limit. It can be seen from
the phase diagram that a wide TSC phase with
Chern number ${\cal N}=1$ emerges between the ${\cal N}=2$ and ${\cal N}=0$ phases, which
are adiabatically connected to the QAH and trivial insulator phases,
respectively.

{\bf Edge picture} An intuitive way to understand such a TSC
phase emerging near the QAH/NI transition is through the
evolution of the edge states. As was discussed earlier, the edge
state of the QAH state (e.g., the point A in the phase diagram in Fig.
\ref{MDphase}(a)) is described by the effective one-dimensional
Hamiltonian $H_{edge}=\sum_{p}vp\eta_{p}^\dagger \eta_{p}$. We can
decompose the complex fermion operator into its real parts,
$\eta_{p_y}=1/\sqrt{2}(\gamma_{p_y 1}+i\gamma_{p_y 2})$
and $\eta^{\dagger}_{p_y}=1/\sqrt{2}(\gamma_{-p_y 1}-i\gamma_{-p_y
2})$ where $\gamma_{p_y a}$ are Majorana fermion
  operators satisfying $\gamma^{\dagger}_{p_y a}=\gamma_{-p_y a}$ and
$\left\{\gamma_{-p_y
a},\gamma_{p^{'}_{y}b}\right\}=\delta_{ab}\delta_{p_{y}p^{'}_{y}}.$
The Hamiltonian now becomes
\begin{equation}
H_{edge}=\sum_{p_y\geq 0}p_{y}\left(\gamma_{-p_{y} 1}\gamma_{p_{y}
1}+\gamma_{-p_{y} 2}\gamma_{p_{y} 2}\right)\end{equation} up to a
trivial shift of the energy. In comparison with the edge theory of
the chiral TSC state, we see that the QAH edge state can be
considered as two identical copies of chiral Majorana fermions, so
that the QAH phase with Chern number $N=1$ can be considered as a
TSC state with Chern number ${\cal N}=2$, {\em even for infinitesimal
pairing amplitudes}. This is consistent with the Chern
number analysis of the bulk Hamiltonian discussed earlier.

\begin{figure}[t!]
\includegraphics[width=0.45\textwidth] {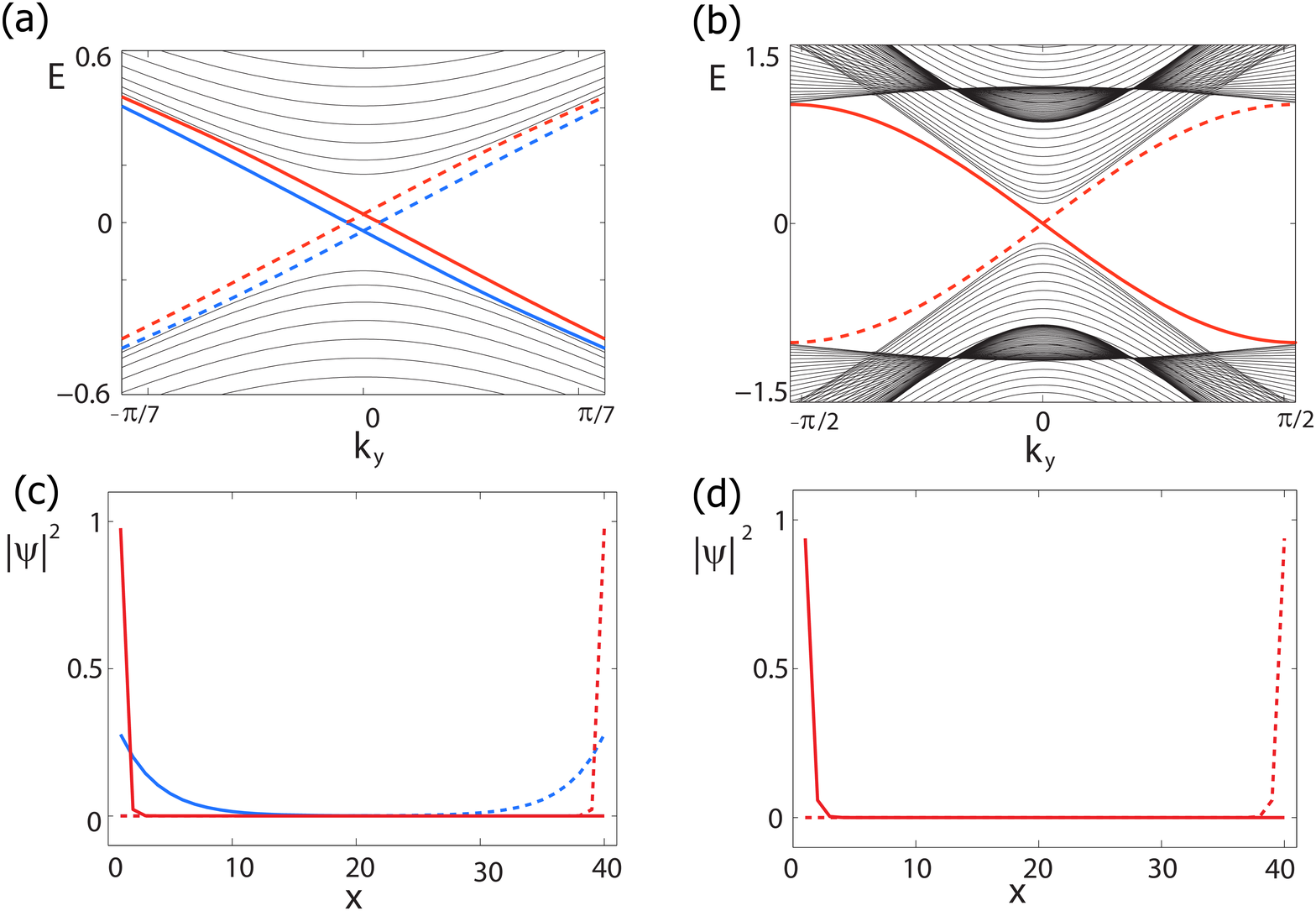}
\caption{Energy spectrum of $H_{BdG}$ versus $k_y$ in a cylinder geometry with periodic boundary conditions in the $y$-direction. The calculation is done for $m=-0.5,A=B=1.0$, with a regularization lattice constant $a=1$ and length $L_x=40$ along the $x$-direction (see text).
The system has two edge states on each edge for $\Delta=0.35$ as shown in (a), and one edge state on each edge for $\Delta=0.75$ as shown in (b). The edge states
on the left (right) edge are labeled by solid (dashed) lines. In (a) the two edge states on the same edge are labeled by red and blue lines. (The two edge
 states are degenerate. An artificially small splitting is added as a guide for the eye) (c) and (d) show the
real space probability distribution of the edge states corresponding to (a) and (b), respectively. The red and blue lines in (c) correspond to the edge states denoted by the same color in (a).
}
\label{fig:edgestate}
\end{figure}

When $\Delta\neq 0$, the constraint between the two chiral Majorana
modes $\gamma_{p_y1}$ and $\gamma_{py2}$ is lifted and they evolve independently. An easy way to see this is
to consider the width of the edge states along the direction
perpendicular to the edge. At $\Delta=0$, the edge state at $k=0$
has a width \cite{koenig2008} $\xi\sim A/|m|$. For finite pairing
their width can be estimated by $\xi_1\sim A/\vert
m-\Delta\vert,\xi_2\sim A/\vert m+ \Delta\vert$.
As $\Delta$ increases, the localization length of one of the edge
modes begins to diverge, and the corresponding Majorana modes
gradually move into the bulk, as shown schematically in Fig.
\ref{fig:edgeschematic} (b). At the critical pairing strength $\vert
\Delta\vert=\vert m\vert$ this edge state has completely merged into
the bulk states and the system is gapless. For $\vert \Delta\vert
>\vert m\vert$ a gap opens again in the bulk,  and
we are left with a single chiral Majorana edge mode. When the mass
is increased toward the trivial direction, such as from point C to D
 in Fig. (\ref{MDphase}) (a), the width of the remaining edge
state increases and finally merges into the bulk at the
critical line, leaving a trivial gapped superconductor on the other
side of the transition.

To directly verify this edge state evolution picture, we have also
studied the edge states numerically on a cylinder geometry with
periodic boundary conditions in the $y$-direction and open boundary
conditions in the $x$-direction. The BdG Hamiltonian (\ref{Hbdg}) can
be regularized on a square lattice by the simple substitution
$p_{x,y}\rightarrow a^{-1}\sin
\left(p_{x,y}a\right),~p^2=2-2a^{-2}\left(\cos p_xa+\cos
p_ya\right)$ with $a$ the lattice constant used in the
discretization. For $a\rightarrow 0$ the lattice model has the same
properties as the continuum model. The energy dispersion and edge
state probability density for the points B and C in the phase diagram
Fig. \ref{MDphase}a are shown in Fig. \ref{fig:edgestate}. As expected, two chiral edge
states with different penetration lengths exist on each edge for the ${\cal N}=2$ phase
and only one chiral edge state exists for the ${\cal N}=1$ phase. Thus we
see that the edge state of the QAH can be considered as two copies of
chiral Majorana fermions. The topological phase transitions are given
by the annihilation of these two Majorana fermions, which has to
occur simultaneously if charge is conserved ($\Delta=0$), but
can generically occur at two separate transitions at finite pairing
strength. Consequently, the system must be in a TSC phase with odd Chern number between the two
transitions.

{\bf Discussion and experimental realization}
In summary, we have shown that the proximity of a QH or a QAH system with an
s-wave superconductor provides a new realization of a chiral
TSC, the vortex of which has non-Abelian
statistics. Although the simple model (\ref{Hqah}) of the $N=1$ QAH state has been
used to give an explicit example, it is straightforward to see that
this approach towards TSC is generic and
applies to any QH or QAH system near a topological phase transition. In
general, a QH or a QAH state with Hall conductance $Ne^2/h$ becomes a
TSC with Chern number ${\cal N}=2N$ when an
{\it infinitesimal} pairing strength is introduced by the proximity
effect. Consequently, two neighboring quantum Hall phases with Hall
conductance $Ne^2/h$ and $(N-1)e^2/h$ become TSC phases with Chern number
${\cal N}=2N$ and ${\cal N}=2N-2$, between
which a TSC phase with Chern number ${\cal N}=2N-1$
generally emerges. This approach is also independent of the details
of the superconductor which provides the proximity effect. As long
as the proximity induced superconducting gap is a full gap without
nodes, our approach is always valid. Our conclusion
also applies to an ordinary quantum Hall system near a plateau
transition, provided the magnetic field responsible for
the QH state is less than the upper critical field of the superconductor.
The QAH system does not require a large magnetic field, which makes the
proximity effect with a superconductor much easier.

There are two realistic proposals of QAH states, ${\rm Mn}$ doped ${\rm HgTe}$
quantum wells\cite{liu2009C}, and Cr or Fe doped ${\rm Bi_2Se_3}$ thin films\cite{yu2010}. The
latter is proposed to be ferromagnetic, which thus can have
quantized Hall conductance at zero magnetic field. The former is
known to be paramagnetic for small Mn concentration, but only a
small magnetic field is needed to polarize the Mn spin and drive the
system into the QAH phase. This is not so prohibitive because a magnetic field is
necessary to generate superconducting vortices and the associated Majorana zero
modes anyway.  Once such
a heterostructure of a QAH insulator and superconductor is fabricated, the
existence of a Majorana fermion zero mode in the vortex core can be
verified by scanning tunneling microscopy (STM) measurements of the
local density of states in the vortex core. Several existing
proposals of detecting the Majorana nature of the edge state and
vortex core zero mode such as by Josephson
effect\cite{fu2009b,lutchyn2010}, charge
transport\cite{akhmerov2009,fu2009} or nonlocal
tunneling\cite{fu2010}, may also apply to our system, although they
are proposed in different physical systems.

{\bf Acknowledgement}. This work is supported by the NSF under grant number DMR-0904264. XLQ acknowledges the support of Microsoft Research Station Q. TLH was supported in part by NSF under grant number DMR-0758462  and by the ICMT.
\bibliography{TI}

\begin{thebibliography}{19}
\expandafter\ifx\csname natexlab\endcsname\relax\def\natexlab#1{#1}\fi
\expandafter\ifx\csname bibnamefont\endcsname\relax
  \def\bibnamefont#1{#1}\fi
\expandafter\ifx\csname bibfnamefont\endcsname\relax
  \def\bibfnamefont#1{#1}\fi
\expandafter\ifx\csname citenamefont\endcsname\relax
  \def\citenamefont#1{#1}\fi
\expandafter\ifx\csname url\endcsname\relax
  \def\url#1{\texttt{#1}}\fi
\expandafter\ifx\csname urlprefix\endcsname\relax\def\urlprefix{URL }\fi
\providecommand{\bibinfo}[2]{#2}
\providecommand{\eprint}[2][]{\url{#2}}

\bibitem[{\citenamefont{Thouless et~al.}(1982)\citenamefont{Thouless, Kohmoto,
  Nightingale, and den Nijs}}]{thouless1982}
\bibinfo{author}{\bibfnamefont{D.~J.} \bibnamefont{Thouless}},
  \bibinfo{author}{\bibfnamefont{M.}~\bibnamefont{Kohmoto}},
  \bibinfo{author}{\bibfnamefont{M.~P.} \bibnamefont{Nightingale}},
  \bibnamefont{and} \bibinfo{author}{\bibfnamefont{M.}~\bibnamefont{den Nijs}},
  \bibinfo{journal}{Phys. Rev. Lett.} \textbf{\bibinfo{volume}{49}},
  \bibinfo{pages}{405} (\bibinfo{year}{1982}).

\bibitem[{\citenamefont{Haldane}(1988)}]{Haldane1988}
\bibinfo{author}{\bibfnamefont{F.~D.~M.} \bibnamefont{Haldane}},
  \bibinfo{journal}{Phys. Rev. Lett.} \textbf{\bibinfo{volume}{61}},
  \bibinfo{pages}{2015} (\bibinfo{year}{1988}).

\bibitem[{\citenamefont{\textrm{X.L. Qi}
  et~al.}(2006)\citenamefont{\textrm{X.L. Qi}, \textrm{Y.S. Wu}, and
  \textrm{S.C. Zhang}}}]{qi2005}
\bibinfo{author}{\bibnamefont{\textrm{X.L. Qi}}},
  \bibinfo{author}{\bibnamefont{\textrm{Y.S. Wu}}}, \bibnamefont{and}
  \bibinfo{author}{\bibnamefont{\textrm{S.C. Zhang}}}, \bibinfo{journal}{Phys.
  Rev. B} \textbf{\bibinfo{volume}{74}}, \bibinfo{pages}{085308}
  (\bibinfo{year}{2006}).

\bibitem[{\citenamefont{Liu et~al.}(2008)\citenamefont{Liu, Qi, Dai, Fang, and
  Zhang}}]{liu2009C}
\bibinfo{author}{\bibfnamefont{C.-X.} \bibnamefont{Liu}},
  \bibinfo{author}{\bibfnamefont{X.-L.} \bibnamefont{Qi}},
  \bibinfo{author}{\bibfnamefont{X.}~\bibnamefont{Dai}},
  \bibinfo{author}{\bibfnamefont{Z.}~\bibnamefont{Fang}}, \bibnamefont{and}
  \bibinfo{author}{\bibfnamefont{S.-C.} \bibnamefont{Zhang}},
  \bibinfo{journal}{Phys. Rev. Lett.} \textbf{\bibinfo{volume}{101}},
  \bibinfo{pages}{146802} (\bibinfo{year}{2008}).

\bibitem[{\citenamefont{Yu et~al.}()\citenamefont{Yu, Zhang, Zhang, Zhang, Dai,
  and Fang}}]{yu2010}
\bibinfo{author}{\bibfnamefont{R.}~\bibnamefont{Yu}},
  \bibinfo{author}{\bibfnamefont{W.}~\bibnamefont{Zhang}},
  \bibinfo{author}{\bibfnamefont{H.~J.} \bibnamefont{Zhang}},
  \bibinfo{author}{\bibfnamefont{S.~C.} \bibnamefont{Zhang}},
  \bibinfo{author}{\bibfnamefont{X.}~\bibnamefont{Dai}}, \bibnamefont{and}
  \bibinfo{author}{\bibfnamefont{Z.}~\bibnamefont{Fang}},
  \bibinfo{howpublished}{arxiv: 1002.0946}.

\bibitem[{\citenamefont{Wilczek}(2009)}]{wilczek2009}
\bibinfo{author}{\bibfnamefont{F.}~\bibnamefont{Wilczek}},
  \bibinfo{journal}{Nat. Phys.} \textbf{\bibinfo{volume}{5}},
  \bibinfo{pages}{614} (\bibinfo{year}{2009}).

\bibitem[{\citenamefont{Volovik}(1999 [JETP Lett. 70, 609
  (1999)])}]{volovik1999}
\bibinfo{author}{\bibfnamefont{G.~E.} \bibnamefont{Volovik}},
  \bibinfo{journal}{Pis'ma Zh. Eksp. Teor. Fiz.} \textbf{\bibinfo{volume}{70}},
  \bibinfo{pages}{601} (\bibinfo{year}{1999 [JETP Lett. 70, 609 (1999)]}).

\bibitem[{\citenamefont{Read and Green}(2000)}]{read2000}
\bibinfo{author}{\bibfnamefont{N.}~\bibnamefont{Read}} \bibnamefont{and}
  \bibinfo{author}{\bibfnamefont{D.}~\bibnamefont{Green}},
  \bibinfo{journal}{Phys. Rev. B} \textbf{\bibinfo{volume}{61}},
  \bibinfo{pages}{10267} (\bibinfo{year}{2000}).

\bibitem[{\citenamefont{Ivanov}(2001)}]{ivanov2001}
\bibinfo{author}{\bibfnamefont{D.~A.} \bibnamefont{Ivanov}},
  \bibinfo{journal}{Phys. Rev. Lett.} \textbf{\bibinfo{volume}{86}},
  \bibinfo{pages}{268} (\bibinfo{year}{2001}).

\bibitem[{\citenamefont{Nayak et~al.}(2008)\citenamefont{Nayak, Simon, Stern,
  Freedman, and Sarma}}]{nayak2008}
\bibinfo{author}{\bibfnamefont{C.}~\bibnamefont{Nayak}},
  \bibinfo{author}{\bibfnamefont{S.~H.} \bibnamefont{Simon}},
  \bibinfo{author}{\bibfnamefont{A.}~\bibnamefont{Stern}},
  \bibinfo{author}{\bibfnamefont{M.}~\bibnamefont{Freedman}}, \bibnamefont{and}
  \bibinfo{author}{\bibfnamefont{S.~D.} \bibnamefont{Sarma}},
  \bibinfo{journal}{Rev. Mod. Phys.} \textbf{\bibinfo{volume}{80}},
  \bibinfo{pages}{1083} (\bibinfo{year}{2008}).

\bibitem[{\citenamefont{Mackenzie and Maeno}(2003)}]{mackenzie2003}
\bibinfo{author}{\bibfnamefont{A.~P.} \bibnamefont{Mackenzie}}
  \bibnamefont{and} \bibinfo{author}{\bibfnamefont{Y.}~\bibnamefont{Maeno}},
  \bibinfo{journal}{Rev. Mod. Phys.} \textbf{\bibinfo{volume}{75}},
  \bibinfo{pages}{657} (\bibinfo{year}{2003}).

\bibitem[{\citenamefont{Fu and Kane}(2008)}]{fu2008}
\bibinfo{author}{\bibfnamefont{L.}~\bibnamefont{Fu}} \bibnamefont{and}
  \bibinfo{author}{\bibfnamefont{C.~L.} \bibnamefont{Kane}},
  \bibinfo{journal}{Phys. Rev. Lett.} \textbf{\bibinfo{volume}{100}},
  \bibinfo{pages}{096407} (\bibinfo{year}{2008}).

\bibitem[{\citenamefont{Sau et~al.}(2010)\citenamefont{Sau, Lutchyn, Tewari,
  and {Das Sarma}}}]{sau2010}
\bibinfo{author}{\bibfnamefont{J.~D.} \bibnamefont{Sau}},
  \bibinfo{author}{\bibfnamefont{R.~M.} \bibnamefont{Lutchyn}},
  \bibinfo{author}{\bibfnamefont{S.}~\bibnamefont{Tewari}}, \bibnamefont{and}
  \bibinfo{author}{\bibfnamefont{S.}~\bibnamefont{{Das Sarma}}},
  \bibinfo{journal}{Phys. Rev. Lett.} \textbf{\bibinfo{volume}{104}},
  \bibinfo{pages}{040502} (\bibinfo{year}{2010}).

\bibitem[{\citenamefont{{K\"{o}nig} et~al.}(2008)\citenamefont{{K\"{o}nig},
  Buhmann, Molenkamp, Hughes, Liu, Qi, and Zhang}}]{koenig2008}
\bibinfo{author}{\bibfnamefont{M.}~\bibnamefont{{K\"{o}nig}}},
  \bibinfo{author}{\bibfnamefont{H.}~\bibnamefont{Buhmann}},
  \bibinfo{author}{\bibfnamefont{L.~W.} \bibnamefont{Molenkamp}},
  \bibinfo{author}{\bibfnamefont{T.}~\bibnamefont{Hughes}},
  \bibinfo{author}{\bibfnamefont{C.-X.} \bibnamefont{Liu}},
  \bibinfo{author}{\bibfnamefont{X.-L.} \bibnamefont{Qi}}, \bibnamefont{and}
  \bibinfo{author}{\bibfnamefont{S.-C.} \bibnamefont{Zhang}},
  \bibinfo{journal}{J. Phys. Soc. Jpn} \textbf{\bibinfo{volume}{77}},
  \bibinfo{pages}{031007} (\bibinfo{year}{2008}).

\bibitem[{\citenamefont{Fu and Kane}(2009{\natexlab{a}})}]{fu2009b}
\bibinfo{author}{\bibfnamefont{L.}~\bibnamefont{Fu}} \bibnamefont{and}
  \bibinfo{author}{\bibfnamefont{C.~L.} \bibnamefont{Kane}},
  \bibinfo{journal}{Phys. Rev. B} \textbf{\bibinfo{volume}{79}},
  \bibinfo{pages}{161408} (\bibinfo{year}{2009}{\natexlab{a}}).

\bibitem[{\citenamefont{Lutchyn et~al.}()\citenamefont{Lutchyn, Sau, and
  Sarma}}]{lutchyn2010}
\bibinfo{author}{\bibfnamefont{R.~M.} \bibnamefont{Lutchyn}},
  \bibinfo{author}{\bibfnamefont{J.~D.} \bibnamefont{Sau}}, \bibnamefont{and}
  \bibinfo{author}{\bibfnamefont{S.~D.} \bibnamefont{Sarma}},
  \bibinfo{howpublished}{e-print arXiv:1002.4033 (2010)}.

\bibitem[{\citenamefont{Akhmerov et~al.}(2009)\citenamefont{Akhmerov, Nilsson,
  and Beenakker}}]{akhmerov2009}
\bibinfo{author}{\bibfnamefont{A.~R.} \bibnamefont{Akhmerov}},
  \bibinfo{author}{\bibfnamefont{J.}~\bibnamefont{Nilsson}}, \bibnamefont{and}
  \bibinfo{author}{\bibfnamefont{C.~W.~J.} \bibnamefont{Beenakker}},
  \bibinfo{journal}{Phys. Rev. Lett.} \textbf{\bibinfo{volume}{102}},
  \bibinfo{pages}{216404} (\bibinfo{year}{2009}).

\bibitem[{\citenamefont{Fu and Kane}(2009{\natexlab{b}})}]{fu2009}
\bibinfo{author}{\bibfnamefont{L.}~\bibnamefont{Fu}} \bibnamefont{and}
  \bibinfo{author}{\bibfnamefont{C.~L.} \bibnamefont{Kane}},
  \bibinfo{journal}{Phys. Rev. Lett.} \textbf{\bibinfo{volume}{102}},
  \bibinfo{pages}{216403} (\bibinfo{year}{2009}{\natexlab{b}}).

\bibitem[{\citenamefont{Fu}(2010)}]{fu2010}
\bibinfo{author}{\bibfnamefont{L.}~\bibnamefont{Fu}}, \bibinfo{journal}{Phys.
  Rev. Lett.} \textbf{\bibinfo{volume}{104}}, \bibinfo{pages}{056402}
  (\bibinfo{year}{2010}).

\end{thebibliography}
\end{document}